\documentclass[conference]{IEEEtran}
\usepackage{verbatim}
\usepackage{graphicx}
\usepackage{amsmath}
\usepackage{amsfonts}

\begin{document}

\title{Obstacles Incentivize Human Learning: A Network Theoretic Study}

\author{
\IEEEauthorblockN{Amitash Ramesh}
\IEEEauthorblockA{Indian Statistical Institute\\
Taramani, Chennai - 600113\\
TN, India\\
amitashr@gmail.com}
\and
\IEEEauthorblockN{Soumya Ramesh}
\IEEEauthorblockA{Indian Statistical Institute\\
Taramani, Chennai - 600113\\
TN, India\\
soumya666@gmail.com}
\and
\IEEEauthorblockN{Sudarshan Iyengar}
\IEEEauthorblockA{Indian Statistical Institute\\
Taramani, Chennai - 600113\\
TN, India\\
sudarshaniisc@gmail.com}

\and
\IEEEauthorblockN{Vinod Sekhar}
\IEEEauthorblockA{Indian Statistical Institute\\
Taramani, Chennai - 600113\\
TN, India\\
vinods1991@gmail.com}
}

\maketitle

\begin{abstract}

The current paper is an investigation towards understanding
the navigational performance of humans on a network when
the `landmark’' nodes are blocked. We observe that humans
 learn to cope up, despite the continued introduction of blockages
in the network. The experiment proposed involves the task
of navigating on a word network based on a puzzle called the 
wordmorph. We introduce blockages in the network and report an 
incremental improvement in performance with respect to time. We 
explain this phenomenon by analyzing the evolution of the  knowledge 
in the human participants of the underlying network as more and more 
landmarks are removed. We hypothesize that humans learn the bare 
essentials to navigate unless we introduce blockages in the network 
which would whence enforce upon them the need to explore newer ways 
of navigating. We draw a parallel to human problem solving and 
postulate that obstacles are catalysts for humans to innovate 
techniques to solve a restricted variant of a familiar 
problem.

\end{abstract}
\IEEEpeerreviewmaketitle

\section{\label{introduction_section}Introduction}

The cognitive problem solving toolkit that humans possess has 
played a pivotal role in shaping the human race in its current form.
From the invention of wheel to the engineering marvels of the information 
age, mankind has evolved tackling problems with his reasoning ability. 
Apart from the ability to solve problems, the greater grandeur lies 
in his invention of machines that can solve problems~\cite{Mainzer07}.

A first study in human problem solving dates back to 1935~\cite{Duncker35} 
at the Berlin School of experimental psychology. The study comprised of 
understanding the strategies used by humans in solving problems in logic. 
Newell in 1958~\cite{Newell58} proposed an architecture for human 
cognition claiming that it can be interpreted as an information processing 
system. It was the pioneering work of Newell and Simon in 
1972~\cite{Simon72} that paved the way for using computers as a tool to 
study human cognition. They proposed the concept of problem solving being 
search in the problem space and asserted that puzzles such as the 
\emph{Tower of Hanoi} have correspondence with problems in real world and 
the study of human strategies to solve such puzzles would open up answers 
for questions on human problem solving in general.

We consider a class of problems whose solution involves starting from a 
premise and arriving at a conclusion. We call such a technique the 
\emph{linear logic strategy}. E.g., The \emph{towers of Hanoi} or the
\emph{Rubik's cube} puzzle admit linear logic strategy. We start from an 
initial state \emph{start} and arrive at a final state \emph{end}, 
using intermediate states $s_1,s_2,...,s_n$. 
$$start\rightarrow s_1 \rightarrow s_2 \rightarrow s_3 \rightarrow s_4 \dots \rightarrow s_n \rightarrow end$$

Several problems in mathematics admit linear logic strategy, we present 
below an illustration of the strategy using a basic algebra problem. Find positive integers 
$a,b,c$ such that, $$abc+3ab+ac+3a+bc+3b+c=27$$
The solution is a chain of steps involving simplification of the given 
expression. Consider,

\begin{equation}\label{eq1}LHS=abc+3ab+ac+3a+bc+3b+c\end{equation} 

\begin{equation}\label{eq2}=ab(c+3)+a(c+3)+b(c+3)+c\end{equation} 

\begin{equation}\label{eq3}=ab(c+3)+a(c+3)+b(c+3)+c+3-3\end{equation}

\begin{equation}\label{eq4}=ab(c+3)+a(c+3)+b(c+3)+1(c+3)-3\end{equation}

\begin{equation}\label{eq5}= (c+3)[ab+a+b+1]-3\end{equation}

\begin{equation}\label{eq6}= (c+3)[a(b+1)+1(b+1)]-3\end{equation}

\begin{equation}\label{eq7}= (c+3)(a+1)(b+1)-3\end{equation}

By inspection one sees that $a=1,b=2,c=2$ yields the required RHS, 27.

Consider the network of all possible equations that one can arrive at starting 
from equation (1). In such a network, equations form the nodes. Two equations 
are considered adjacent if one can be obtained from the other by a 
\emph{unit step operation}, which involves introducing brackets and performing 
algebraic decomposition or adding and subtracting an element. With only one such 
operation permitted at every step, we observe that solving this problem 
translates to navigating on the network of algebraic expressions.\\
\\ \noindent
\emph{\bf Introducing Obstacles:} If one is restricted from using equation 
(\ref{eq2}), that is, to not add and subtract an integer, will one come out with 
an entirely different strategy to solve this problem? The main theme of this 
paper would be to answer this question in a framework that would help in the 
better conduction and analysis of the experiment. \\

\subsection{Previous work}
A study on how humans navigate on an unfamiliar network was first conducted by 
Sudarshan et. al. in~\cite{sudarshan11}. Their study involved human participants 
to navigate from a source node to a destination node on a network. The authors 
present an analysis of the paths that were taken by 20 human participants where 
each were asked to solve 50 node pairs. They note that humans exhibit a tendency
to recognize \emph{landmarks} in the network and are inclined to using them as 
via-media to navigate from a given source node to a destination node. They show 
that these landmark words are nodes with superior centrality ranking (for details 
on centrality, refer~\cite{zweig05}). They also observe that the paths that 
humans take are seldom optimal and that they never attempt to learn more than 
what is required to navigate on the network. They consider a word puzzle called 
the \emph{wordmorph}, where one is asked to find a path from a source word to 
the destination word by changing one slot at a time. E.g., given the word pair 
$(TOY,KID)$ a valid solution would be $TOY-TON-TAN-PAN-PAT-SAT-SIT-KIT-KID$. 
The rules of the game permit the usage of valid English dictionary words 
disallowing proper nouns (E.g. USA). We provide details of the experiment in 
Section.~\ref{experiment_section}. Wordmorph game was introduced by Lewis 
Carroll~\cite{Gardner96} of the `Alice in Wonderland' fame on a Christmas day in 
1877. A first analysis of this network was reported by Knuth in 
1993~\cite{Knuth93}.
  
\subsection{Present work}
By a \emph{wordmorph network} we mean the network of three lettered words with a 
link between every two words that differ in precisely one slot. As noted in the 
work of Sudarshan et. al.~\cite{sudarshan11}, humans learn a few landmark words 
and use them repeatedly in order to navigate on the network. This is analogous 
to a real life situation, where a traveling saleswoman\footnote{Not to be 
confused with the optimization problem popularly known as the Traveling salesman problem} 
who is new to the city is required to navigate between any two given locations. 
Eventually, the saleswoman will recognize and learn the landmarks in the city 
and use them to reach the desired destination. 
\emph{
\begin{quotation}
Once our participants recognize and start using the landmark words, we progressively
start removing them. This creates an \underline{obstacle} in our participants' progress and challenges
them to explore new landmarks and better ways to navigate on the disrupted network. 
We \underline{incentivize} our participants to explore the network by introducing obstacles.
\end{quotation}
}
In an interesting experiment conducted by Moeser~\cite{Moeser88}, nurses in an 
hospital with a complicated building structure, never managed to learn better 
routes to navigate. They learnt the bare minimum ways inside the building and 
used the same to navigate despite their stay in the same place for 2 consecutive 
years. This is well explained by Passini who proposed in~\cite{Passini84} that, 
when humans are asked to way-find, they learn what is necessary and sufficient 
to achieve the goal and never improvise on it. 

Present work is an extension of the wordmorph experiment as reported 
in~\cite{sudarshan11} (We explain the previous work in detail in 
Section. \ref{experiment_section}.) \\

We broadly address the following questions:

\begin{enumerate}

\item Does one get complacent with a familiar technique that one doesn't 
doesn't make an attempt to find alternate ways of solving the same problem?

\item Does one learn an altogether new strategy when one is faced with 
obstacles/restrictions? 

\item Does one learn more than what is required to surpass the obstacle?
\end{enumerate}

We present related work in Section~\ref{literature_section} followed by 
preliminaries and definitions in Section~\ref{preliminaries_section}. The details 
of the experiment is presented in Section~\ref{experiment_section} followed by 
the results and discussions in Section~\ref{results_section}. We end with our 
concluding remarks in Section~\ref{conclusion_section}.

\begin{comment}
Consider the following trigonometric identity :

$sin^2x+cos^2x=1$

If one is asked to prove the above identity, one will resort to the following proof:
    
Given a right angled triange $ABC$, with $AC$ being the hypotenuse, 
let $x = \angle ACB$ then,\\ $sin(x)=\frac{AB}{AC}$ and $cos(x)=\frac{BC}{AC}$.\\
We know by \emph{Pythagoras Theorem} that \\
$AC^2=AB^2+BC^2$\\ $\Rightarrow$ $\frac{AB}{AC}^2 +\frac{BC}{AC}^2=1$.

In case one is asked to prove the above identity without the usage 
of the Pythagoras theorem, then one is forced to try a novel strategy. E.g.
a student with a knowledge of first course in calculus will readily 
deploy the following method:

Consider $f(x)=sin^2x+cos^2x$, \\
$f'(x)=2sinxcosx-2cosxsinx =0$\\ 
$Rightarrow f(x)$ is a constant function.\\
$f(0)=1$, $\Rightarrow f(x)=sin^2x+cos^2x=1, \forall x$
 
\end{comment}

\section{\label{literature_section}Literature}

\subsection{Networks}In the recent years, a type of complex environment, called the complex network 
has interested researchers across several disciplines. It is known that most 
real-world networks belong to the type of so called small worlds~\cite{Watts98}, 
that is, they have a small average distance and is thus easy to reach each node 
in the network within only a few steps. This was first demonstrated by a classic 
experiment by~\cite{Milgram67} in which he asked participants to send a letter 
via acquaintances to a person unknown to them. Most of the letters reached their 
target within only a few steps. This finding was reproduced by Watts 
et. al.~\cite{Dodds03} using e-mails. Kleinberg was the first to ask and address 
the question as to how people are essentially able to find short paths in a small 
world without any geometric information~\cite{Kleinberg00}. As there are 
potentially many paths of short length, this is indeed not totally obvious. 

\subsection{Human Navigation}
Human navigation in a complex environment has been a topic studied extensively 
under spatial cognition from past few decades~\cite{McDonald93,Moore76}. 

Understanding and analyzing the learning involved in wayfinding has very important implications 
in explaining spatial behaviour. When moving around in a large environment, humans learn (visual)
landmarks which help them to navigate from a source to a destination. There has been a significant 
number of studies in human spatial cognition and development~\cite{Aginsky97,Evans80,McDonald93}. 
It has been observed that humans have an inherent tendency to learn to navigate from one place to the other by remembering 
spatial landmarks.

Of special interest to us is the work by Moeser~\cite{Moeser88}, in which he reports on results of a
study conducted in a 5-storey hospital which had a very complicated structure. It was noted
that the student nurses learnt a set of landmarks in the hospital network for navigation and 
did not learn new ways to navigate better even after 2 years of their stay in the hospital as 
there was no need for the same. If the experiment had been conducted with obstacles induced in 
the most frequently used paths, we suspect that they may have explored newer and possibly better 
ways to reach the destination.

Aginsky et. al.~\cite{Aginsky97} proposed two strategies that humans adopt to navigate, 
they infer that humans follow either a visually dominated or spatially dominated strategy to 
solve a route-learning problem.

Way-finding is the problem of reaching a specified destination 
from a given source in an unfamiliar environment. 
Basakaya et. al.~\cite{Baskaya04} found that signage placed at decision points of complex buildings improved 
wayfinding performance. Peponis et. al.~\cite{Peponis90} worked on a 
hospital with a complex layout and redirected the attention from an exclusive focus on local 
characteristics, such as signs and landmarks, to one that also considers the overall structure 
of the building. According to Siegel and White~\cite{Siegel75}, 
levels of cognitive mapping begin with landmark elements. They proposed that landmark knowledge 
precedes route knowledge, and both precede configurational knowledge in environmental development. 
Landmarks and/or a zone with a strong character may favor a certain spatial identification in 
the sense of being somewhere distinct. 

An interesting work involving animal navigation, Pavel~\cite{Pavel03} conducted an experiment on 
the wood mouse (Apodemus sylvaticus) observing the way they navigate in a given environment. The 
experiment was conducted by scattering white discs around the experiment environment. It was found 
that the mice positioned these white discs and used them as landmarks to aid in navigation.

Passini~\cite{Passini84} has proposed that in a 
way-finding problem, one learns what is necessary and sufficient to achieve a
goal. Landmarks are essential parts of wayfinding cues and are seen as 
points of reference. Therefore, sensitivity to landmark quality is a critical 
factor for wayfinding as people use landmarks as reference point at necessary 
times. Landmarks are the most prominent cues in any environment~\cite{Darken02}. 
Landmarks act as key elements to enhance the ability to orient oneself and to navigate 
in an environment. their importance is because of aiding the user in navigating and 
understanding the spaces~\cite{Sorrows99}. To mark an object as a landmark among 
the others is done by the individual. 

%#########################################

\subsection{Wordmorph}
To investigate how humans solve the wayfinding problem, Sudarshan et. al.~\cite{sudarshan11} 
proposed an experiment based on a game called the wordmorph. Their study involves 
investigating the strategies that humans adapt in order to navigate in a complex 
network. The wordmorph game presents a well defined navigation problem in a 
complex network: given two words $(w_1, w_2)$ of the same length, for example 
(BOY,PER), one is asked to find a sequence of words from $w_1$ to $w_2$ such that 
each successor differs in only one letter from it's predecessor. Fig. 1 shows 
examples of feasible solutions for the word pairs (CAR,SHY), (AXE,NUT) and 
(TRY,POT). Note that these are just one of the many possible solutions. The
navigation strategy cannot be described by any of the models sketched earlier. 
However, this simple setting allows to  identify the strategy by which people 
learn to navigate. It was observed, that humans quickly identify the so-called
landmark words which they frequently use in their navigation. It is also observed 
that these landmark nodes are centrally located in the complex network, leading 
to a direct correlation between network structure and human navigation in it.

\begin{figure}[h]
\begin{center}
\includegraphics[width=.5\textwidth]{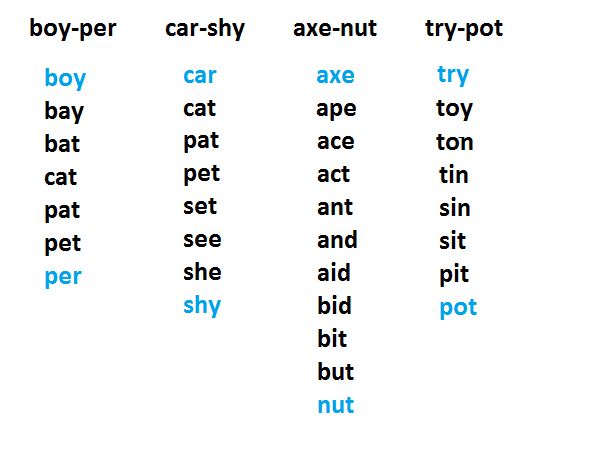}
\caption { Exemplary wordmorph games }
\end{center}
\end{figure}

\section{\label{preliminaries_section} Preliminaries}

\begin{figure}[h]
\begin{center}
\includegraphics[width=.5\textwidth]{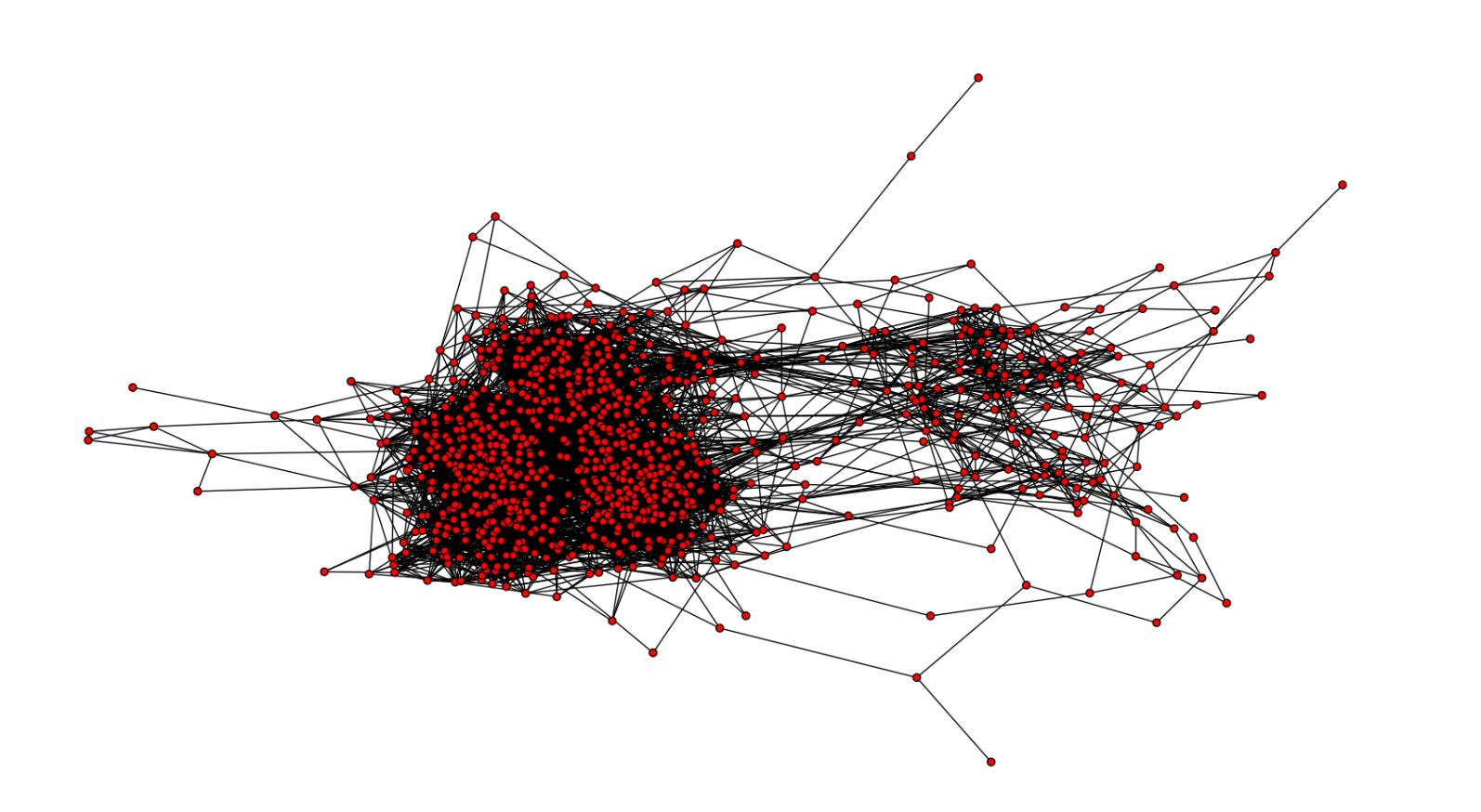}
\caption {The wordmorph network  $G(L_3)$ composed of all three-letter words in English.}
\label{amitash}

\end{center}
\end{figure}

A graph $G=(V,E)$ is composed of a set of nodes $V$ and a set of edges 
$E\subseteq V\times V$, with $|V|=n$ and $|E|=m$. A {\it way} between two nodes 
$u$ and $v$ is any sequence of edges $(e_1, e_2, \dots, e_k)$ with 
$e_1=(u,x_1), e_2=(x_1,x_2), \dots ,e_k=(x_{k-1},v)$. A {\it path} is a way 
with no repeating nodes. The length of a way is defined as the number of edges in it. 
A \emph{shortest path} between two nodes $u$ and $v$ is a path with length $l$, such that, 
every other path between $u$ and $v$ is of length greater than or equal to $l$.
Two nodes are said to be {\it connected} if there exists a path between them. The 
{\it distance} $d(v_1, v_2)$ between any two nodes is defined as the length 
of a shortest path between them, or set to $\infty$ if there exists no path 
between them. Any maximal set of pairwise connected nodes is 
called a {\it component} of the graph. 
Let $\Sigma$ be a set of letters, and $\Sigma^*$ be the set of all possible 
concatenations. Let  $L\subseteq \Sigma^*$ be some language and $L_k$ denote 
the set of all words with the same number of letters $k$.  For a given $L$ and $k$, 
one instance of a {\it wordmorph game} consists of two words 
$(start, end) \in L_k\times L_k$. A {\it solution} of this game is any sequence 
of words $start=w_1, w_2, w_3, \dots, w_k=end$ such that any two consecutive words 
differ in exactly one letter. E.g., for the pair $(CAR, SHY)$, the following 
sequence is a solution: 
\begin{center}$(CAR, CAT, PAT, PET, SET, SEE, SHE, SHY)$.\end{center}
The rules of the wordmorph game defines a natural relation $\simeq_R$ on all words 
in $L_k$, i.e., regarding the rules, any two words $v,w$ in $L_k$ are related if 
they differ by exactly one letter. Thus, $(L_k,\simeq_R)$ defines a graph on the 
words in $L_k$, which we call the {\it wordmorph network} $G(L_k)$ on $L_k$. In the 
following, $L_k$ will be the set of all three-letter words in English, as defined 
by the Oxford dictionary~\cite{Oxford05}, and $G(L_3)$ is the respective graph. $G(L_3)$ is
shown in Fig.\ref{amitash}.

%\begin{comment}
%\end{comment}

A {\it centrality index} is a real-valued function $C: V \rightarrow \mathbb R$ 
on the nodes~\cite{zweig05}. The intuition is that the higher the value of this 
function, the more central this node is for the network. There are various 
indices; in this article we use the so-called {\it closeness centrality} 
~\cite{Sabidussi66} $C_C(v)$, which is defined as the reciprocal of sum of the distances of $v$ to all other nodes $w$: 
$$C_C(v) = 1/\sum_{w\in V} d(v,w)$$
For any given graph, a centrality index can be used to define a ranking on the nodes~\cite{sudarshan11}. We use this 
centrality measure to rank the nodes in a players wordmorph network to define a difficulty parameter during the conduction of our present experiment.

To understand the exploration and navigation of humans in a complex network with obstacles, we conducted a series of wordmorph 
games with $15$ different participants. The experimental setting will be described in the following section.

\section{\label{experiment_section}The Experiment}

\bigskip
\noindent
{\bf Analysis from Previous Experiments:} Our experiment is an extension of experiments conducted in~\cite{sudarshan11}. 
Participants in the previous work were made to navigate on a wordmorph network. 
It was observed that humans learn to navigate efficiently using certain key words 
or "Landmark Words". In addition, it was seen that the paths taken by humans are 
highly center strategic in nature. A path is said to be center strategic if 
participants use landmark words to navigate. These landmark words were further 
seen to be of high closeness centrality in nature. In our current experiment, 
we observe human navigational performance when these landmark words are blocked.
 
\bigskip\noindent
{\bf The Experimental Setting:}

The experiment was conducted on 15 participants (8 men and 7 women). The 
participants whose data are presented in our analysis are all those who completed 
around 255 games except for one participant who dropped out after a few games. 
On an average, each participant took around 7 hours to finish the game. The data 
for the 3250 games are analyzed. 9 participants were graduate students of the age 
group 20-25 and the remaining 6 were working adults in the age group of 25-35. 
None of them knew the game beforehand.

\begin{figure}[h]
\begin{center}
\includegraphics[width=.5\textwidth]{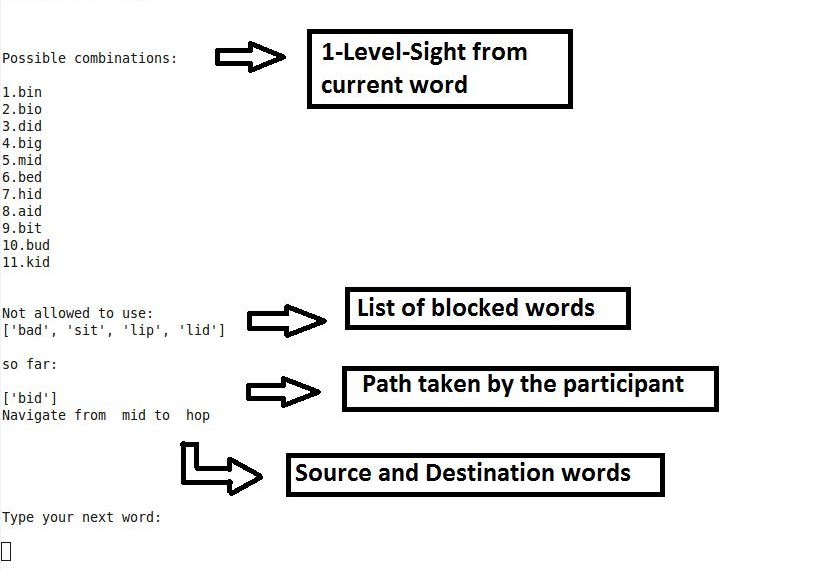}
\caption { Snapshot of the Game Interface}
\label{snapshot}
\end{center}
\end{figure}

From the list of all 967 three-letter words contained in the Oxford English 
Dictionary ~\cite{Oxford05}, each participant Pi first selected the words she 
knew. This set is denoted by $V(P_i)$ and the respective graph is denoted by $G(P_i)$. 
The game was explained to the participant and we informed her that the winning strategy 
is to identify landmarks as the interest in our experiment was to remove these 
landmarks and analyze their performance. In our experiment we provided the users 
with a "1-level-sight", i.e., we displayed all the possible words she could navigate 
to, from the current word. After the blockage of landmark words, these blocked words 
were displayed to the participant and she was informed not to use them anymore. 
A snapshot of the game interface is provided in Fig.\ref{snapshot}. Out of the 15 
participants, 5 participants were made to play without obstacles and their results 
are analyzed as well.

\bigskip \noindent
{\bf Creating the Wordmorph instances:}

A sample set of 55 games were given to the participants out of which the first 15 
games were given so that they were made familiar with the playing interface. All 
(source,destination) word pairs were chosen to be distinct. The difficulty for the 
remaining 40 games was set by giving them word pairs of low closeness centrality 
(bottom 30\%) and the minimum distance between the word pairs. $d(w_i,w_j)$ was 5. 
This was easily possible, as the diameter which is defined as 
$max[d(u,v):u,v \in V]$ of the graphs G(Pi) on average was 12.3. At the end of the 
sample set, the 4 most frequently used words by the participant, i.e., the initial 
set of 4 landmarks were blocked. This difficulty with respect to path length and 
low closeness centrality ranking was maintained henceforth throughout the game.
The remaining 200 games were divided into 5 phases with phase 1 comprising of games 
1-40, phase 2 comprising of games 41-80, phase 3 comprising of games 81-120, phase 
4 comprising of games 121-160 and phase 5 comprising of games 161-200. Phase 1 begins 
with the removal of the 4 most frequently used words in the sample set. Henceforth, 
at every phase, the 4 most frequently used words in the previous phase were removed 
(punched) from the graph.We chose to remove landmarks every 40 games, as it is seen 
from previous experiments~\cite{sudarshan11} that 40 games is a very sufficient 
condition for a participant to learn landmarks. It was also seen from those experiments 
that on an average, a participant uses 4 words with a much greater frequency than 
any other words.

\bigskip\noindent
{\bf Information Logging and Post-Processing of Data:}

For each participant Pi, her selected vocabulary V(Pi) was saved and G(Pi) computed.
 For this graph, the closeness centrality was computed for all nodes and their rank was
determined by sorting the words accordingly. For each of the 255 game instances, we 
stored the word pairs that was given to the participant and her solution. Time was 
recorded for all 15 participants. Participants entered their solutions via an 
interactive computer program. They were not allowed to use any writing aids.

\bigskip\noindent
{\bf Follow up Experiment Without Obstacles:}

5 Participants amongst the 15 were made to play 250 games of the same difficulty parameters 
as before. However, we did NOT block their landmark words. Their games were also recorded 
and the final analysis is explained in the following section.

%\begin{comment}
%\end{comment}

\section{\label{results_section}Results and Discussions}

Based on thorough analysis of data obtained from our experiments, 
we find several interesting phenomena. We observe that humans 
cope with obstacles in way finding very effectively and give an 
explanation for their improved performance. We also compare our 
results to navigation without obstacles which gives us an insight 
into knowledge acquisition by humans. We summarize our findings 
based on the following hypotheses.

\bigskip\noindent
{\bf 1. Humans adapt quickly and efficiently to successive obstacles.}

On removal of landmark words for the first time, a struggle is 
observed which is captured in the large increase in time taken 
by the participants to navigate on the "disrupted" network.
Participants find it hard to use center strategic paths and hence
the increase in time is observed. As the participants continue 
navigating, we observe that they learn a new set of landmark
words which is seen when they restore to the usage of center
 strategic paths. We remove these words again and observe the 
time taken in the  second phase. The average time taken in this
 phase shows that the struggle in this phase is lesser compared
 to that in the  previous phase. We progressively remove the 
landmark words after equal interval of games thus increasing the difficulty. 
At the end of the game, we asked each of the participants the 
following question:

\emph{\quotation How difficult was the game on each blockage?}\\

\begin{figure}[h]
\begin{center}
\includegraphics[width=.4\textwidth]{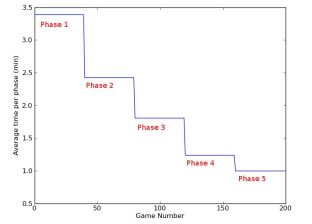}
\caption { Average Time per Phase for a Single Participant}
\label{time}
\end{center}
\end{figure}

All the participants answered that after an initial struggle, 
they were able to navigate easily even on further route 
blockages. To substantiate this, we observe that the average 
time taken per phase decreases further in successive phases. 
We can see from Fig.\ref{time} and Table I   that towards the end, the participants 
are very comfortable and navigate very quickly. Even though it 
may seem that humans are increasingly hindered by progressively 
introducing obstacles, we observe that: In fact, humans get 
better at handling these impediments, and eventually overcome them by 
finding new landmark nodes. This indicates that humans learn to adapt to
obstacles quickly!

An intriguing question that arises is why this happens and what 
exactly are humans learning to overcome these obstacles?

\begin{table}[ht]
\caption{Time Taken per Phase} % title of Table
\centering  % used for centering table
\begin{tabular}{c  c  c  c  c  c } % centered columns (4 columns)
\hline\hline                        %inserts double horizontal lines
Participant & Phase 1 & Phase 2 & Phase 3 & Phase 4 & Phase 5  \\ [1.0ex] % inserts table 
%heading
\hline                  % inserts single horizontal line

1 & 3.38 & 2.42 & 1.80 & 1.23 & 0.99\\
2 & 2.38 & 2.29 & 2.12 & 1.97 & 1.89\\
3 & 2.71 & 2.43 & 2.25 & 1.90 & 1.70\\
4 & 1.42 & 1.21 & 1.07 & 0.95 & 0.75\\
5 & 2.42 & 1.86 & 1.25 & 1.08 & 0.94\\
6 & 1.55 & 1.39 & 1.24 & 0.95 & 0.66\\
7 & 2.21 & 1.90 & 1.80 & 1.31 & 1.11\\
8 & 3.29 & 2.74 & 1.98 & 1.47 & 1.05\\
9 & 3.44 & 2.85 & 2.25 & 1.91 & 1.54\\

\hline %inserts single line
\end{tabular}
\label{table:nonlin3} % is used to refer this table in the text
\end{table}

%\newpage

%\begin{comment}
%\end{comment}
{\bf 2: Obstacles Induce Learning:}

We consider that a node is \emph{'Learnt'} when the participant visits it for the first 
time. Similarly we consider an edge to be learnt when a participant traverses across it 
for the first time.

\begin{figure}[h]
\begin{center}
\includegraphics[width=.4\textwidth]{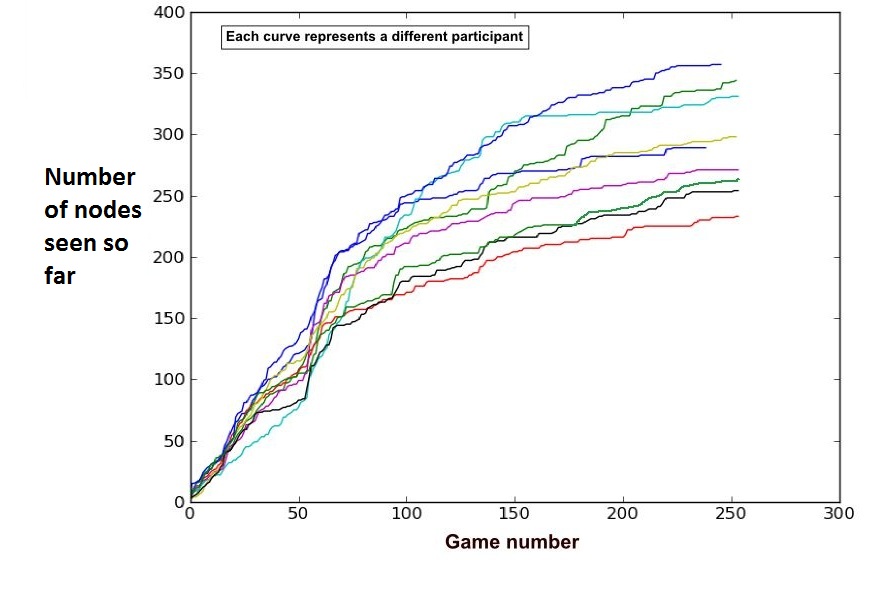}
\caption { Total Nodes Learnt so far Across all Participants}
\label{nodes}
\end{center}
\end{figure}

\begin{figure}[h]
\begin{center}
\includegraphics[width=.4\textwidth]{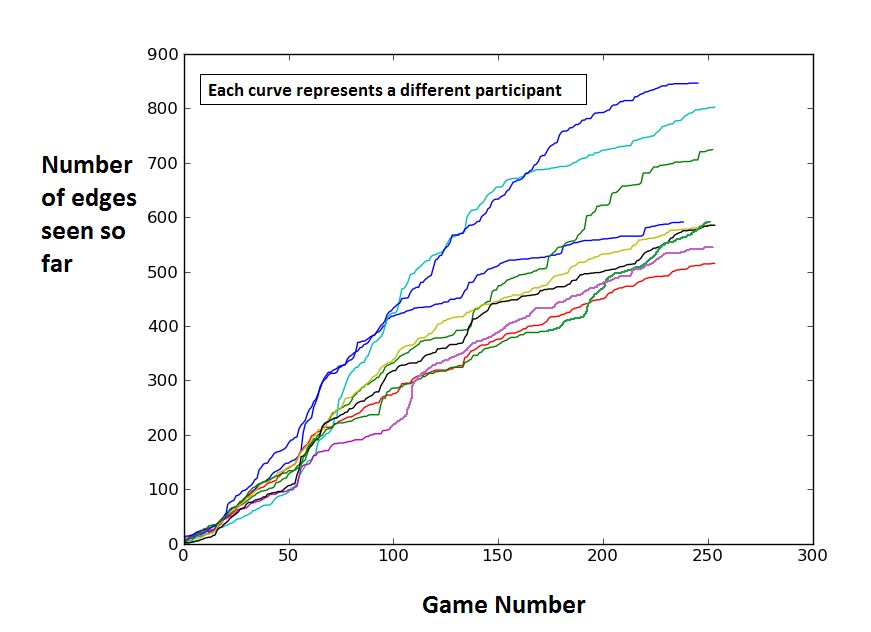}
\caption {Total Edges Learnt so far Across all Participants }
\label{edges}
\end{center}
\end{figure}

We observe that introducing a blockage makes a participant explore newer nodes and edges.
Constant introduction of blockages results in the player learning a large number of nodes 
and edges as seen in Fig.\ref{nodes} and Fig.\ref{edges}. In fact, we observe that by the end of the game,
the participant learns more than 70\% of the network. This increase in a participant's 
initial knowledge base facilitates better navigability in the network. The participant 
becomes adept at quickly finding center strategic paths. Thus, even though a participant
is constantly inhibited with obstacles, his increased knowledge base in the previous games 
enables him to navigate quickly and efficiently within the network. This explains the 
decrease in time over consecutive phases. It would be interesting to see if the 
participants learn the same amount if they were not faced with blockages.
%\begin{comment}
%\end{comment}

%\newpage
%\begin{comment}
%\end{comment}

{\bf 3: Learning is Significantly Lesser if No Obstacles Were Introduced:}

We conducted the same experiment on a different set of participants with a key distinction: 
No obstacles were introduced.

We observe that the players learn a set of landmark words and continue to navigate using 
this existing knowledge throughout the entire 250 games. The learning is drastically lesser
as compared to learning with obstacles as they do not increase their knowledge by exploring 
the network much further. The participants learn to use center strategic paths and continue 
to use them without experiencing any degree of difficulty. Their landmark words remain more 
or less the same throughout all the games. In fact, even when there were shorter paths
available, the participants tend to use longer paths which pass through their landmark
words. Results are tabulated in Table II and Table III

\begin{table}[ht]
\caption{Way Finding With Obstacles} % title of Table
\centering  % used for centering table
\begin{tabular}{c  c  c} % centered columns (4 columns)
\hline\hline                        %inserts double horizontal lines
Participant & Percentage of Nodes Learnt & Percentage of Edges Learnt  \\ [1.0ex] % inserts table 
%heading
\hline                  % inserts single horizontal line

1 &	74.4	&	48.0	\\
2 & 	75.6	&	43.8 	\\
3 & 	74.9	& 	43.4	\\	
4 &	75.2	&	41.6	\\
5 &	79.7	&	46.6	\\
6 &	69.7	& 	39.2	\\
7 &	72.5	&	42.1	\\
8 &	68.6	&	38.2	\\	
9 &	70.3	&	40.8	\\  

\hline %inserts single line
\end{tabular}
\label{table:nonlin1} % is used to refer this table in the text
\end{table}

Average node learning: 73.43\%

Average edge learning: 42.63\%
\begin{table}[ht]
\caption{Way Finding Without Obstacles} % title of Table
\centering  % used for centering table
\begin{tabular}{c c  c} % centered columns (4 columns)
\hline\hline                        %inserts double horizontal lines
Participant & Percentage of Nodes Learnt & Percentage of Edges Learnt  \\ [1.0ex] % inserts table 
%heading
\hline                  % inserts single horizontal line

1 &    39.6    &    18.7  \\ 
2 &    41.5    &    20.4  \\
3 &    42.3    &    21.1  \\
4 &    38.9    &    17.8  \\
5 &    36.5    &    15.9  \\

\hline %inserts single line
\end{tabular}
\label{table:nonlin0} % is used to refer this table in the text
\end{table}

Average node Learning: 39.76\%

Average Edge Learning: 18.78\%

\bigskip
It is intuitive that blockages force the participant to learn more of the network, but how 
much more do they learn and is it really necessary for them to learn so much? This is 
analyzed in the next section.
 
\bigskip
{\bf 4: Humans Learn Much More than What is Required on Introduction of Obstacles:}

\begin{figure}[h]
\begin{center}
\includegraphics[width=.5\textwidth]{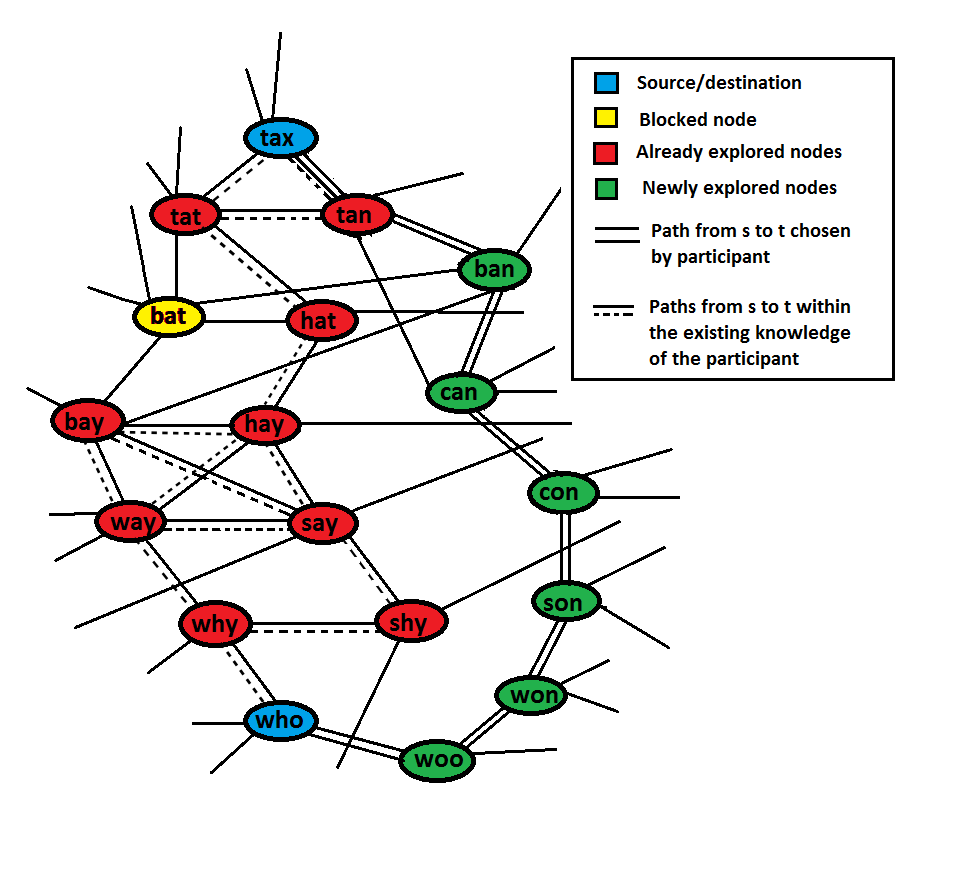}
\caption { Extraneous Learning }
\label{extra}
\end{center}
\end{figure}

\begin{figure}[h]
\begin{center}
\includegraphics[width=.5\textwidth]{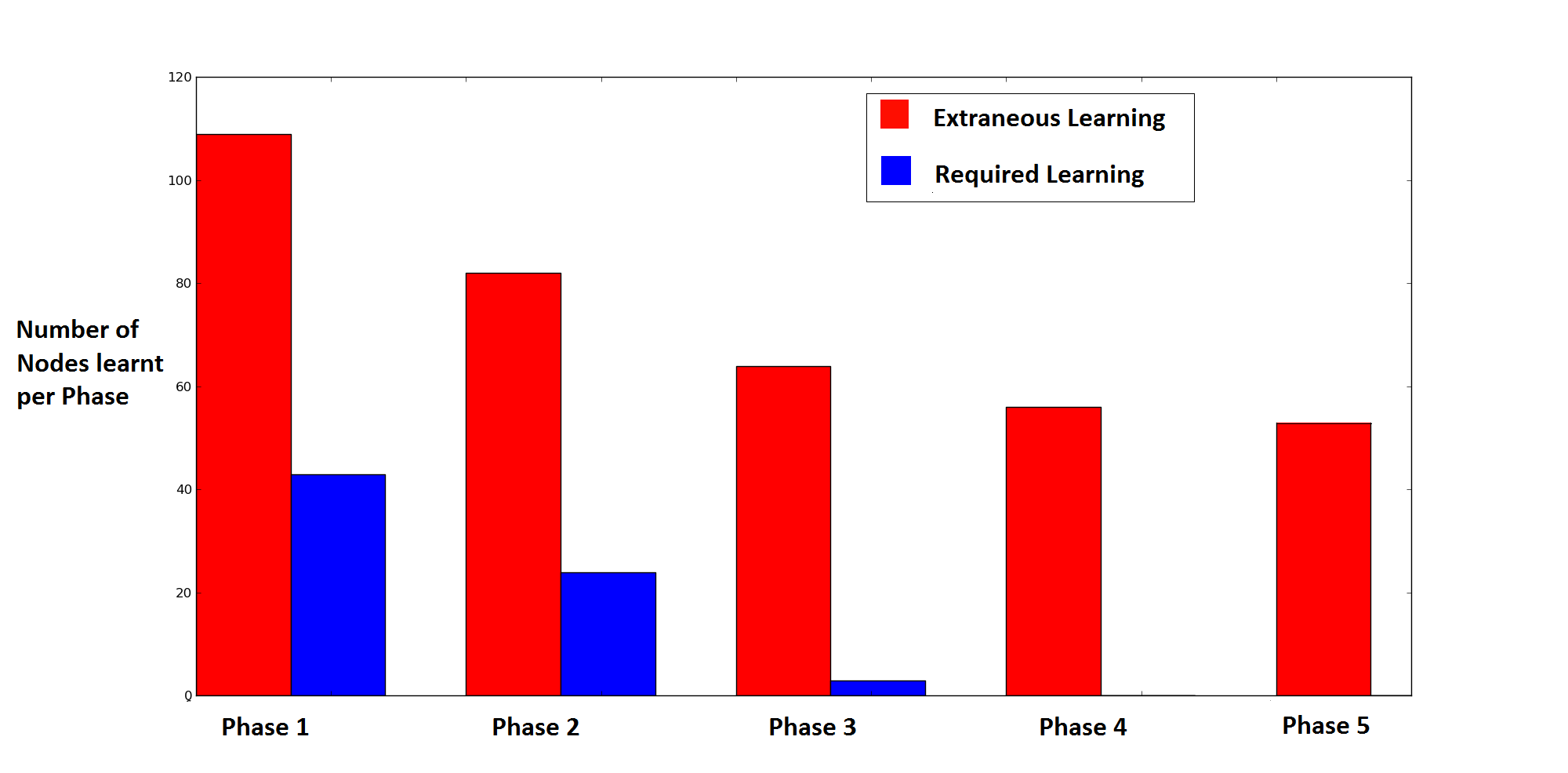}
\caption { Nodes Learnt vs Required Learning per Phase for a Single Participant}
\label{single}
\end{center}
\end{figure}

When landmark words are removed, it is expected that participants tend to learn new words 
to navigate through the network. However, when we calculate the amount of words that would 
have been sufficient for navigation on the disrupted network, we observe that participants 
undergo an extraneous amount of learning when they are faced with an obstacle. i.e: While 
the participant could have found a quick route by finding several other paths within his 
knowledge base, he actually takes a longer and more circuitous route, thereby increasing 
his knowledge base (Fig.\ref{extra}). Fig.\ref{single} Shows learning comparison for a single participant. 
We calculate the necessary number of words required per game by finding the shortest path 
from the source and destination words to the existing knowledge of the participants. 
We find that humans tend to move away from their knowledge base and explore a lot more of 
the network even though their existing knowledge was enough for them to navigate in. 
The results are tabulated as shown in Table IV.

{\bf Thus, on introduction of obstacles, humans tend to increase their learning by an amount 
that was rather unnecessary.}

%\begin{comment}
%\end{comment}

%\begin{comment}
%\end{comment}

%\newpage
\begin{table}[ht]
\caption{Number of nodes required Vs Number of nodes learnt} % title of Table
\centering  % used for centering table
\begin{tabular}{c  c  c } % centered columns (4 columns)
\hline\hline                        %inserts double horizontal lines
Participant & Required Learning & Actual Learning   \\ [1.0ex] % inserts table 
%heading
\hline                  % inserts single horizontal line

1&  56 & 279\\
2&  48 & 328\\ 
3&  70 & 381\\
4&  66 & 347\\
5&  59 & 302\\
6&  71 & 323\\
7&  70 & 373\\
8&  65 & 316\\
9&  71 & 327\\
\hline %inserts single line
\end{tabular}
\label{table:nonlin2} % is used to refer this table in the text
\end{table}

\section{\label{conclusion_section}Conclusion and Further Work}

This network theoretic study aims at understanding human navigation 
in a network with progressive disruptions. We analyze how humans 
explore such a network and the resulting learning that transpires.
 We presented an analysis of the wordgame whose solution involved linear logic strategy. As proposed by ~\cite{Simon72}, a friendly and easy framework to understand human problem solving is by understanding the strategies of humans in solving problems such as the Tower of Hanoi and puzzles that involve the linear logic strategy. Wordmorph puzzle facilitates such a study with a 2 folded advantage, one being the network is in the order of hundreds of nodes thus making it non-trivial an exercise and the other being the humans familiarity of the wordmorph network which has enabled us to conduct this experiment on people of diverse age groups and experience.
 Based on the observables from our experiment, we make a generalized hypotheses about human problem solving:

{\bf 1.} We note that humans exhibit exploratory skills when they are faced with obstacles in problem solving.

{\bf 2.} Not only do they succeed in learning novel ways of navigating through the problem space,they 'learn to learn quickly' with time.

{\bf 3.} The most startling observation is the amount of extraneous learning that humans undergo in order to overcome obstacles.Apart from man's cognitive capabilities being the primary reason for
the enormous amount of knowledge acquisition that mankind has undergone during the course of human evolution, it seems a strong secondary reason lies in its learning ability of extrinsic information when he is fraught with obstacles.

A very interesting further study would be to validate our hypothesis by considering a group of high school students who have undergone a course in axiomatic geometry. Students are known to solve problems in geometry with a few axioms and postulates as their bag of tricks\footnote{History has it that the problems known for its notoriety like the Fermat's last Theorem have yielded an entirely new school of thought in mathematics and have paved a long way in the understanding of various other facets of the subject~\cite{Fermatenigma}}. Will they explore better ways of solving when they are asked  not to use a particular geometry factual? Such a study would require a good interdisciplinary approach across topics such as mathematical problem solving, learning psychology, network analysis and the like.

\section{\label{ack}Acknowledgements}

We would like to thank the Indian Statistical Institute for their gracious support and valuable funding for our experiment.
We would also like to thank Dayananda Sagar College of Engineering - Department of Computer Science for their support as well.
And lastly, we thank all our participants who spent a lot of their time and effort with us.

\bibliography{references}

\bibliographystyle{IEEEtran}

\end{document}